%% file: main.tex
\documentclass[runningheads,a4paper]{llncs}
\usepackage[top=2.5cm, bottom=2.5cm, left=2.5cm, right=2.5cm]{geometry}
\usepackage{amsmath}
\usepackage{amssymb}
\usepackage{graphicx}
\usepackage{booktabs}
\usepackage{listings}
\usepackage{xcolor}
\usepackage{hyperref}
\usepackage{caption}
\usepackage{subcaption}

\usepackage{url}
\usepackage{color}
\usepackage{cite}
\usepackage{epsfig}

\usepackage[nomargin,inline,marginclue,draft]{fixme}
\usepackage{cleveref}
\fxsetup{status=draft}
\fxsetup{theme=color, mode=multiuser} \FXRegisterAuthor{me}{ame}{\color{red}Me}
\newcommand{\orcid}[1]{\href{https://orcid.org/#1}{\includegraphics[scale=0.02]{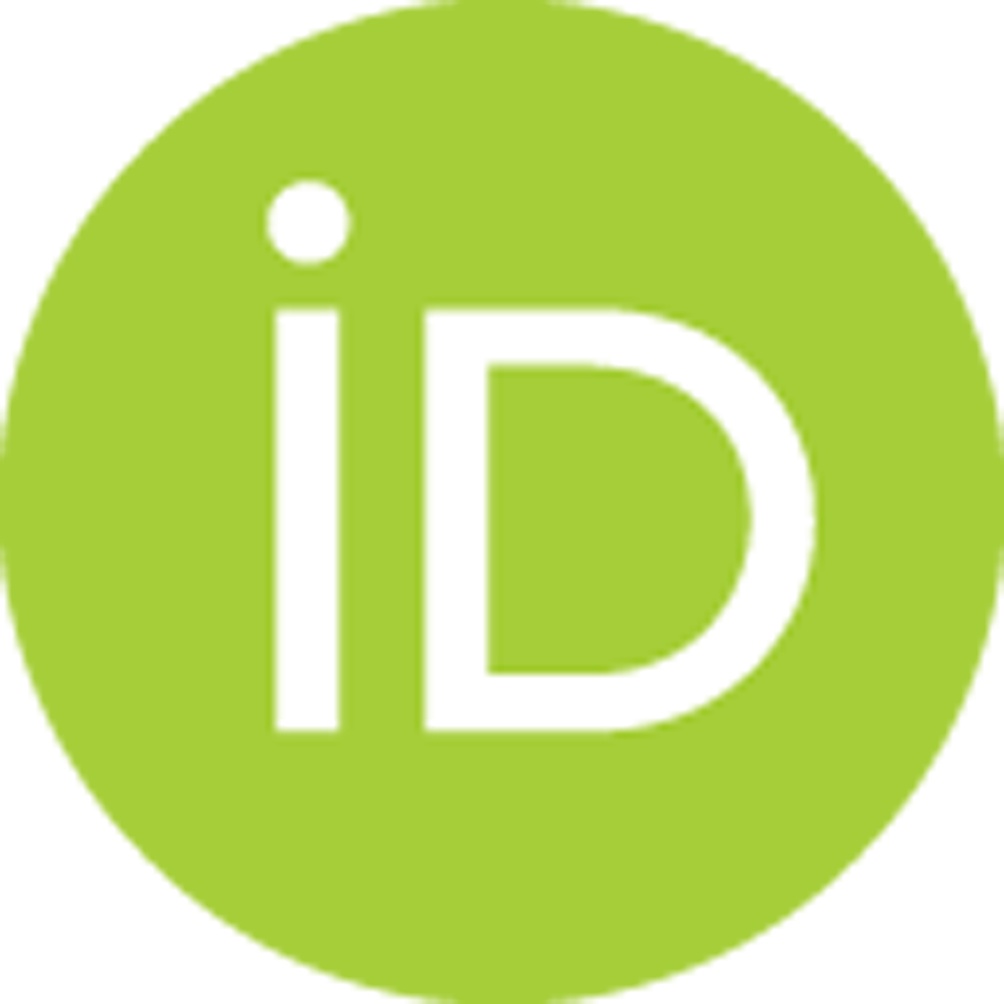}}} 

\hypersetup{
    colorlinks=true
}
\title{The Brain Tumor Segmentation (BraTS) Challenge 2023: \emph{Focus on Pediatrics (CBTN-CONNECT-DIPGR-ASNR-MICCAI BraTS-PEDs)}}
\titlerunning{The BraTS-PEDs 2023 Challenge}

\input{0_author_list}

\begin{document}
    \mainmatter
    \maketitle
    \setcounter{footnote}{0} 
     \begin{abstract}
         Pediatric tumors of the central nervous system are the most common cause of cancer-related death in children. The five-year survival rate for high-grade gliomas in children is less than 20\%. Due to their rarity, the diagnosis of these entities is often delayed, their treatment is mainly based on historic treatment concepts, and clinical trials require multi-institutional collaborations. The MICCAI Brain Tumor Segmentation (BraTS) Challenge is a landmark community benchmark event with a successful history of 12 years of resource creation for the segmentation and analysis of adult glioma. Here we present the CBTN-CONNECT-DIPGR-ASNR-MICCAI BraTS-PEDs 2023 challenge, which represents the first BraTS challenge focused on pediatric brain tumors with data acquired across multiple international consortia dedicated to pediatric neuro-oncology and clinical trials. 
         The BraTS-PEDs 2023 challenge focuses on benchmarking the development of volumentric segmentation algorithms for pediatric brain glioma through standardized quantitative performance evaluation metrics utilized across the BraTS 2023 cluster of challenges. Models gaining knowledge from the BraTS-PEDs multi-parametric structural MRI (mpMRI) training data will be evaluated on separate validation and unseen test mpMRI dataof high-grade pediatric glioma. The CBTN-CONNECT-DIPGR-ASNR-MICCAI BraTS-PEDs 2023 challenge brings together clinicians and AI/imaging scientists to lead to faster development of automated segmentation techniques that could benefit clinical trials, and ultimately the care of children with brain tumors.
    \end{abstract}
    
    
    \input{1_introduction}
         
    \section{Materials \& Methods}
        \input{2_Material_and_Methods.tex}

        \input{3_Results}
        \input{4_discussion}

    \section*{Acknowledgments}
        Success of any challenge in the medical domain depends upon the quality of well annotated multi-institutional datasets. We are grateful to all the data contributors, annotators, and approvers for their time and efforts. Our profound thanks go to the Children's Brain Tumor Network (CBTN), the Collaborative Network for Neuro-oncology Clinical Trials (CONNECT), the International DIPG/DMG Registry (DIPGR), the American Society of Neuroradiology (ASNR), and the Medical Image Computing and Computer Assisted Intervention (MICCAI) Society for their invaluable support of this challenge. 
    
    \section*{Funding}
    
    Research reported in this publication was partly supported by the National Institutes of Health (NIH) under award numbers: NCI/ITCR:U01CA242871 and NCI:UH3CA236536, and by grant funding from the Pediatric Brain Tumor Foundation and DIPG/DMG Research Funding Alliance (DDRFA).

    \bibliographystyle{ieeetr}
    \bibliography{bibliography.bib}
    \newpage
    \appendix
\end{document}

%% file: 0_author_list.tex
\author{
Anahita Fathi Kazerooni \inst{1,2,3,\dag,\ddag,\orcid{0000-0001-7131-2261}}
\and Nastaran Khalili \inst{1,\dag, \S, *,\orcid{0000-0001-8078-3591}}
\and Xinyang Liu \inst{4,\dag,*}
\and Debanjan Haldar \inst{1,5,\dag, \S, *}
\and Zhifan Jiang \inst{4,\dag}
\and Syed Muhammed Anwar \inst{4, \dag}
\and Jake Albrecht \inst{6, \dag}
\and Maruf Adewole \inst{7,\dag}
\and Udunna Anazodo \inst{8,\dag}
\and Hannah Anderson \inst{1, \S}
\and Sina Bagheri \inst{1, \S}
\and Ujjwal Baid \inst{3,9,10,\dag,\orcid{0000-0001-5246-2088}}
\and Timothy Bergquist \inst{6,\dag}
\and Austin J. Borja \inst{11, \S}
\and Evan Calabrese \inst{12,\dag}
\and Verena Chung \inst{6,\dag}
\and Gian-Marco Conte \inst{13,\dag}
\and Farouk Dako \inst{14,\dag}
\and James Eddy \inst{6,\dag}
\and Ivan Ezhov \inst{15,16,\dag}
\and Ariana Familiar \inst{1, \ddag}
\and Keyvan Farahani \inst{17,\dag}
\and Shuvanjan Haldar \inst{1,18, \S}
\and Juan Eugenio Iglesias \inst{19,\dag}
\and Anastasia Janas \inst{20,\dag}
\and Elaine Johansen \inst{21,\dag}
\and Blaise V Jones \inst{22, \S, \orcid{0000-0001-9003-4578}}
\and Florian Kofler \inst{23,\dag}
\and Dominic LaBella \inst{24,\dag}
\and Hollie Anne Lai \inst{25, \S, \orcid{0000-0001-6747-7338}}
\and Koen Van Leemput \inst{26,\dag}
\and Hongwei Bran Li \inst{19,\dag}
\and Nazanin Maleki \inst{20,\ddag}
\and Aaron S McAllister \inst{27, \S, \orcid{0000-0002-1117-9355}}
\and Zeke Meier \inst{28,\dag}
\and Bjoern Menze \inst{29,30,\dag}
\and Ahmed W Moawad \inst{31,\dag}
\and Khanak K Nandolia \inst{32, \S, \orcid{0000-0002-8627-2992}}
\and Julija Pavaine \inst{33, 34, \S, \orcid{0000-0003-1843-837X}}
\and Marie Piraud \inst{23,\dag}
\and Tina Poussaint \inst{35, \ddag}
\and Sanjay P Prabhu \inst{36, \S, \orcid{0000-0003-0871-115X}}
\and Zachary Reitman \inst{24,\dag}
\and Andres Rodriguez \inst{37, \S, \orcid{0000-0002-5366-5351}}
\and Jeffrey D Rudie \inst{38,\dag,\orcid{0000-0000-0000-0000}}
\and Ibraheem Salman Shaikh \inst{39, \S, \orcid{0000-0002-5994-997}} 
\and Lubdha M. Shah \inst{40, \S, \orcid{0000-0003-1303-3533}} 
\and Nakul Sheth \inst{41, \S} 
\and Russel Taki Shinohara \inst{42,\dag}
\and Wenxin Tu \inst{43,\S} 
\and Karthik Viswanathan \inst{1} 
\and Chunhao Wang \inst{24,\dag}
\and Jeffrey B Ware \inst{9,\S}
\and Benedikt Wiestler \inst{30,\dag}
\and Walter Wiggins \inst{12,\dag}
\and Anna Zapaishchykova \inst{35, \ddag}
\and Mariam Aboian \inst{20,\dag, \ddag,\orcid{0000-0002-4877-8271}}
\and Miriam Bornhorst \inst{44,\ddag,\orcid{0000-0000-0000-0000}} 
\and Peter de Blank \inst{45, \ddag} 
\and Michelle Deutsch \inst{46,\ddag}
\and Maryam Fouladi \inst{46,\ddag}
\and Lindsey Hoffman \inst{47}
\and Benjamin Kann \inst{35, \ddag} 
\and Margot Lazow \inst{46,\ddag}
\and Leonie Mikael \inst{46,\ddag}
\and Ali Nabavizadeh \inst{1,9,\dag,\ddag,\S,\orcid{0000-0002-0380-4552}}
\and Roger Packer \inst{44,\ddag}
\and Adam Resnick \inst{1,2,\ddag,\orcid{0000-0003-0436-4189}}
\and Brian Rood \inst{48,\ddag}
\and Arastoo Vossough \inst{1,41,\dag,\S, \orcid{0000-0003-1346-427X}}
\and Spyridon Bakas \inst{3,9,10,\dag,\orcid{0000-0001-8734-6482}}
\and Marius George Linguraru \inst{4,49,\dag,\ddag,**, \orcid{0000-0001-6175-8665}}
}

\authorrunning{Kazerooni A, et al.}

\institute{\scriptsize{ Center for Data-Driven Discovery in Biomedicine (D3b), Children's Hospital of Philadelphia, Philadelphia, PA, USA
\and Department of Neurosurgery, University of Pennsylvania, Philadelphia, PA, USA
\and Center for AI \& Data Science for Integrated Diagnostics (AI2D) and Center for Biomedical Image Computing and Analytics (CBICA), University of Pennsylvania, Philadelphia, PA, USA
\and Sheikh Zayed Institute for Pediatric Surgical Innovation, Children's National Hospital, Washington DC, USA
\and Department of Neurosurgery, Thomas Jefferson University Hospital, PA, USA
\and Sage Bionetworks, USA
\and Medical Artificial Intelligence (MAI) Lab, Crestview Radiology, Lagos, Nigeria
\and Montreal Neurological Institute (MNI), McGill University, Montreal, QC, Canada
\and Department of Radiology, Perelman School of Medicine, University of Pennsylvania, Philadelphia, PA, USA
\and Department of Pathology and Laboratory Medicine, Perelman School of Medicine, University of Pennsylvania, Philadelphia, PA, USA
\and Department of Neurosurgery at the University of Southern California, CA, USA
\and Department of Radiology, Duke University Medical Center, USA
\and Mayo Clinic, MN, USA
\and Center for Global Health, Perelman School of Medicine, University of Pennsylvania, Philadelphia, PA, USA
\and Department of Informatics, Technical University Munich, Germany
\and TranslaTUM - Central Institute for Translational Cancer Research, Technical University of Munich, Germany
\and Cancer Imaging Program, National Cancer Institute, National Institutes of Health, Bethesda, MD, USA
\and Biomedical Engineering Rutgers University, New Brunswick, NJ, USA
\and Athinoula A Martinos Center for Biomedical Imaging, Massachusetts General Hospital, Boston, MA, USA
\and Yale University, New Haven, CT, USA
\and PrecisionFDA, U.S. Food and Drug Administration, Silver Spring, MD, USA
\and Cincinnati Children’s Hospital Medical Center (ORCID ID: 0000-0001-9003-4578
\and Helmholtz AI, Helmholtz Munich, Germany
\and Department of Radiation Oncology, Duke University Medical Center, USA
\and Department of Radiology, Children’s Health Orange County, CA, USA (ORCID: 0000-0001-6747-7338)
\and Department of Applied Mathematics and Computer Science, Technical University of Denmark, Denmark
\and Department of Radiology, Nationwide Children's Hospital, OH, (ORCID: 0000-0002-1117-9355)
\and Booz Allen Hamilton, McLean, VA, USA
\and Biomedical Image Analysis \& Machine Learning, Department of Quantitative Biomedicine, University of Zurich, Switzerland
\and Department of Neuroradiology, Technical University of Munich, Munich, Germany
\and Mercy Catholic Medical Center, Darby, PA, USA
\and Department of Diagnostic and Interventional Radiology, All India Institute of Medical Sciences, Rishikesh, India (ORCID: 0000-0002-8627-2992)
\and Department of Paediatric Radiology, Royal Manchester Children’s Hospital, Manchester University Hospitals NHS Foundation Trust, University of Manchester, Manchester, UK (ORCID: 0000-0003-1843-837X)
\and Division of Informatics, Imaging \& Data Sciences, School of Health Sciences, Faculty of Biology, Medicine and Health, University of Manchester, Manchester Academic Health Science Centre, Manchester, UK
\and Dana-Farber Brigham Cancer Center \& Boston Children’s Hospital, Boston, MA, USA
\and Division of Neuroradiology, Department of Radiology, Boston Children's Hospital, Harvard Medical School, Boston, Massachusetts, USA  (ORCID: 0000-0003-0871-115X)
\and Department of Diagnostic \& Interventional Imaging, Neuroradiology section. The University of Texas Health Sciences Center at Houston. (ORCID: 0000-0002-5366-5351)
\and University of California, San Diego, CA, USA
\and Beth Israel Deaconess Medical Center, Harvard Medical School, MA, USA (ORCID: 0000-0002-5994-997)
\and University of Utah, UT, USA (ORCID: 0000-0003-1303-3533)
\and Department of Radiology, The Children's Hospital of Philadelphia, PA, USA
\and Center for Clinical Epidemiology and Biostatistics, University of Pennsylvania, Philadelphia, USA
\and College of Arts and Sciences, University of Pennsylvania, PA, USA (ORCID: 0000-0003-3649-2994)
\and Brain Tumor Institute, Children's National Hospital, Washington, DC, USA
\and Brain Tumor Center, Cincinnati Children's Hospital, Cincinnati, OH, USA
\and Pediatric Neuro-Oncology Program, Nationwide Children’s Hospital, Columbus, OH, US
\and Center for Cancer and Blood Disorders, Phoenix Children's Hospital, Phoenix, AZ, US
\and Center for Cancer and Blood Disorders, Children’s National Hospital, Washington, DC, USA
\and Departments of Radiology and Pediatrics, George Washington University School of Medicine and Health Sciences, Washington, DC, USA
}
\linebreak
\\
\textsuperscript{\dag} People involved in the organization of the challenge.\\
\textsuperscript{\ddag} People contributing data from their institutions.\\
\textsuperscript{\S} People involved in annotation process.\\ 
\textsuperscript{*} Equal authors.\\
\textsuperscript{**} Corresponding author: \email{\{mlingura@childrensnational.org\}}}

%% file: 1_introduction.tex
\section{Introduction}

   Although rare, pediatric tumors of the central nervous system are the most common cause of cancer-related death in children. While pediatric brain tumors may share certain similarities with adult brain tumors, their imaging and clinical presentations differ. For example, adult glioblastomas (GBMs), pediatric diffuse midline gliomas (DMGs, including pediatric diffuse intrinsic pontine glioma (DIPGs)) are high grade gliomas with short average overall survival \cite{mackay2017integrated, jansen2015survival}. The incidence of GBMs is 3 in 100,000 people, while DMGs are about three times more rare. While GBMs are usually found in the frontal or/and temporal lobes at an average age of 64 years, many DMGs are located in the pons and often diagnosed between 5 and 10 years of age. Enhancing tumor regions on post-gadolinium T1-weighted MRI and radiologically-defined necrotic tissue regions are common imaging findings in GBM, but less common or clear in DMGs, at least at initial diagnosis. Thus, pediatric brain tumors require dedicated imaging tools that help in their characterization and facilitate their diagnosis/prognosis \cite{fathi2023automated,madhogarhia2022radiomics,nabavizadeh2023current}. 
   
   Tumor segmentation is essential in surgical and treatment planning, as well as in response assessment and longitudinal monitoring. However, manual segmentation is time-consuming and has high inter-operator variability \cite{bakas2016glistrboost}. Pediatric brain tumors are variable in their aggressiveness, prognosis, and heterogeneous histologic subregions, i.e., peritumoral edematous/invaded tissue, necrotic core, and enhancing tumor core, as reflected in their radio-phenotypes on multi-parametric magnetic resonance imaging (mpMRI) scans. Some pediatric brain tumors, such as DIPGs, are heterogeneous and typically located in surgically inaccessible anatomical locations, leaving them unresectable. As such, assessment of tumor progression is mainly based on the changes in size of the tumor measured from longitudinal scans. The current standard, as recommended by the Response Assessment in Pediatric Neuro-Oncology (RAPNO) cooperative working group \cite{cooney2020response,erker2020response}, is based on two-dimensional (2D) linear measurements of the tumor subregions (in the axial slice of the largest tumor extent), including enhancing tumor, non-enhancing tumor, cystic components, and peritumoral edematous tissue. Nonetheless, these 2D measurements are inaccurate surrogates of the volume and they cannot accurately account for irregularity of the tumor shape and appearance and hence further increase the inter-operator variability. Studies in adult brain tumors have indicated superiority of 3D volumetric measurement of tumor extent in the prediction of clinical endpoints over 2D methods \cite{ellingson2022volumetric}.  
   While the importance of volumetric tumor measurements, as compared to 2D ones, are being increasingly acknowledged in response assessment of pediatric brain tumors \cite{lazow2022volumetric,jansen2015survival}, there is a lack of available automated tools to facilitate the cumbersome task of segmentation of tumor subregions on associated scans. There are only a handful of automated tumor segmentation methods explicitly proposed for pediatric brain tumors \cite{fathi2023automated, nalepa2022segmenting, artzi2020automatic, tor2020unsupervised, mansoor2016deep, avery2016optic, mansoor2017joint, peng2022deep}. However, the majority of these methods have been developed only for the segmentation of the T2 fluid attenuated inversion recovery (FLAIR) abnormal signal \cite{nalepa2022segmenting,artzi2020automatic,peng2022deep}, also called whole tumor (WT) in the BraTS terminology. Furthermore, none of the proposed methods has been evaluated and compared on the same validation data and through a controlled study design, and therefore there is a gap in benchmarking the automated segmentation tools in pediatric brain tumors.
   
    The MICCAI brain tumor segmentation (BraTS) challenges have established a community benchmark dataset and environment for adult glioma over the past 11 years  \cite{menze2014multimodal,bakas2017advancing,bakas2018identifying,baid2021rsna}. The focus of this year’s BraTS is expanded to a Cluster of Challenges spanning across various tumor entities, missing data, and technical considerations. During BraTS 2022, we organized the first initiative to include pediatric brain tumors, specifically DMGs, in the test phase of the BraTS 2022 challenge. The findings of this evaluation study encouraged us to organize a larger and more diverse initiative in 2023 with multi-institutional pediatric data. In the BraTS-PEDs 2023 challenge, the focus is on pediatric brain tumor entities. We have created a retrospective multi-institutional (multi-consortium) pediatric database, with the data collected through a few consortia, including Children’s Brain Tumor Network (CBTN, \url{https://cbtn.org/} ) \cite{lilly2023children}, DIPG Registry (\url{https://www.dipgregistry.org}) and the COllaborative Network for NEuro-oncology Clinical Trials (CONNECT, \url{https://connectconsortium.org/}). Additional data from participating pediatric institutions have been included in the BraTS-PEDs cohort. The American Society of Neuroradiology (ASNR, \url{https://www.asnr.org/}) collaborated in generating ground truth annotation for the majority of data in this challenge. These data will be collectively used to develop and quantitatively evaluate algorithms developed and models trained by the challenge participants in unseen validation data until July 2023, when the organizers will stop accepting new submissions and evaluate submitted containerized algorithms in a completely unseen pediatric patient population. This manuscript provides details about the CBTN-CONNECT-ASNR-MICCAI BraTS-PEDs 2023 challenge, as the first benchmarking initiative on pediatric brain tumor segmentation.

%% file: 2_Material_and_Methods.tex
\subsection{Data}

    The BraTS-PEDs 2023 dataset includes a retrospective multi-institutional cohort of conventional/structural magnetic resonance imaging (MRI) sequences, including pre- and post-gadolinium T1-weighted (labeled as T1 and T1CE), T2-weighted (T2), and T2-weighted fluid attenuated inversion recovery (T2-FLAIR) images, from 228 pediatric high-grade glioma. These conventional multiparametric MRI (mpMRI) sequences are commonly acquired as part of standard clinical imaging for brain tumors. However, the image acquisition protocols and MRI equipment differ across different institutions, resulting in heterogeneity in image quality in the provided cohort. Inclusion criteria comprised of pediatric subjects with: (1) histologically-approved high-grade glioma, i.e., high-grade astrocytoma and diffuse midline glioma (DMG), including radiologically or histologically-proven diffuse intrinsic pontine glioma (DIPG); (2) availability of all four structural mpMRI sequences on treatment-naive imaging sessions. Exclusion criteria consisted of: (1) images assessed to be of low quality or with artifacts that would not allow for reliable tumor segmentation; and (2) infants younger than one month of age. Data for n = 228 patients was obtained through CBTN (n = 138), Boston's Children Hospital (n = 61), and Yale University (n = 29). 
     
    The cohort included in the BraTS-PEDs 2023 challenge is split into training (n = 99), validation (n = 45), and testing datasets. The data shared with the participants comprise mpMRI scans and ground truth labels for the training cohort, as well as mpMRI sequences without any associated ground truth for the validation cohort. Notably, the testing data that will be used for evaluating the performance of the methods submitted by challenge participants will not be shared with the participants, but the containerized submissions of the participants will be evaluated by the synapse.org platform, powered by MedPerf\cite{karargyris2021medperf}. 
     
    Participants are prohibited from training their algorithm on any additional public and/or private data (from their own institutions) besides the provided BraTS-PEDs 2023 data, or using models pretrained on any other dataset. This restriction is imposed to allow for a fair comparison among the participating methods. However, participants can use additional public and/or private data only for publication of their scientific papers, if they also provide a report of their results on the data from BraTS-PEDs 2023 challenge alone, and discuss potential differences in the obtained results.

\label{sec:data}

    \subsubsection{Imaging Data Description}  

    For all patients, mpMRI scans were pre-processed using a standardized approach, including conversion of the DICOM files to the NIfTI file format \cite{nifti}, co-registration to the same anatomical template (SRI24) \cite{SRI_rohlfing2010sri24}, and resampling to an isotropic resolution ($1mm^{3}$). The pre-processing pipeline is publicly available through the Cancer Imaging Phenomics Toolkit (CaPTk) \footnote{\url{https://cbica.github.io/CaPTk/}} \cite{captk,captk_2,captk_3} and Federated Tumor Segmentation (FeTS) tool \footnote{\url{https://github.com/FETS-AI/Front-End/}}. De-identification was performed through removing protected Health Information (PHI) from DICOM headers in DICOM to NIfTI conversion step \cite{NEJMc1908881,NEJMc1915674}. De-facing was performed via skull-stripping using a formerly proposed pediatric-specific skull-stripping method \cite{fathi2023automated} to prevent any potential facial reconstruction/recognition of the patients.
         
    The resulted images were segmented into tumor subregions using an integration of two pediatric automated deep learning segmentation models \cite{fathi2023automated, liu2023}. The result of this step was segmentation of the tumors into four main subregions, recommended by RAPNO working group for evaluation of the treatment response in high-grade gliomas and DIPGs. The annotated tumor subregions comprised of enhancing tumor (ET), nonenhancing tumor (NET), cystic component (CC), and peritumoral edema (ED) regions. ET is described by areas with enhancement (brightness) on T1 post-contrast images as compared to T1 pre-contrast. In case of mild enhancement, checking the signal intensity of normal brain structure can be helpful. CC typically appears with hyperintense signal (very bright) on T2 and hypointense signal (dark) on T1CE. The cystic portion should be within the tumor, either centrally or peripherally (as compared to ED which is peritumoral). The brightness of CC is here defined as comparable or close to cerebrospinal fluid (CSF). NET is defined as any other abnormal signal intensity within the tumorous region that cannot be defined as enhancing or cystic. For example, the abnormal signal intensity on T1, T2-FLAIR, and T2 that is not enhancing on T1CE should be considered as nonenhancing portion. ED is defined by the abnormal hyperintense signal (very bright) on FLAIR scans. ED is finger-like spreading that preserves underlying brain structure and surrounds the tumor. The automatically-generated four labels (Figure 1) using the automated segmentation tool were used as preliminary segmentation to be manually revised by volunteer neuroradiology experts of varying rank and experience, in accordance with annotation guidelines.  
    The volunteer neuroradiology expert annotators were provided with four mpMRI sequences (T1, T1CE, T2, FLAIR) along with the fused automated segmentation volume to initiate the manual refinements. The ITK-SNAP \cite{itksnap} software was used for making these refinements. Once the automated segmentations were refined by the annotators, three  attending board-certified neuroradiologists, reviewed the segmentations. Depending upon correctness, these segmentations were either approved or returned to the individual annotator for further refinements. This process was followed iteratively until the approvers found the refined tumor subregion segmentations acceptable for public release and the challenge conduction.
    
    While four tumor subregions were annotated, in keeping with the other BraTS 2023 challenges and to allow the participants to incorporate data from adult glioma challenge for their model training, the participants are provided with three segmentation labels, including enhancing tumor (ET), nonenhancing component (NC - a combination of nonenhancing tumor, cystic component, and necrosis), as well as peritumoral edematous area (ED) (Figure 1).

    \begin{figure}[t]
          \centering
          \includegraphics[width=0.9\linewidth]{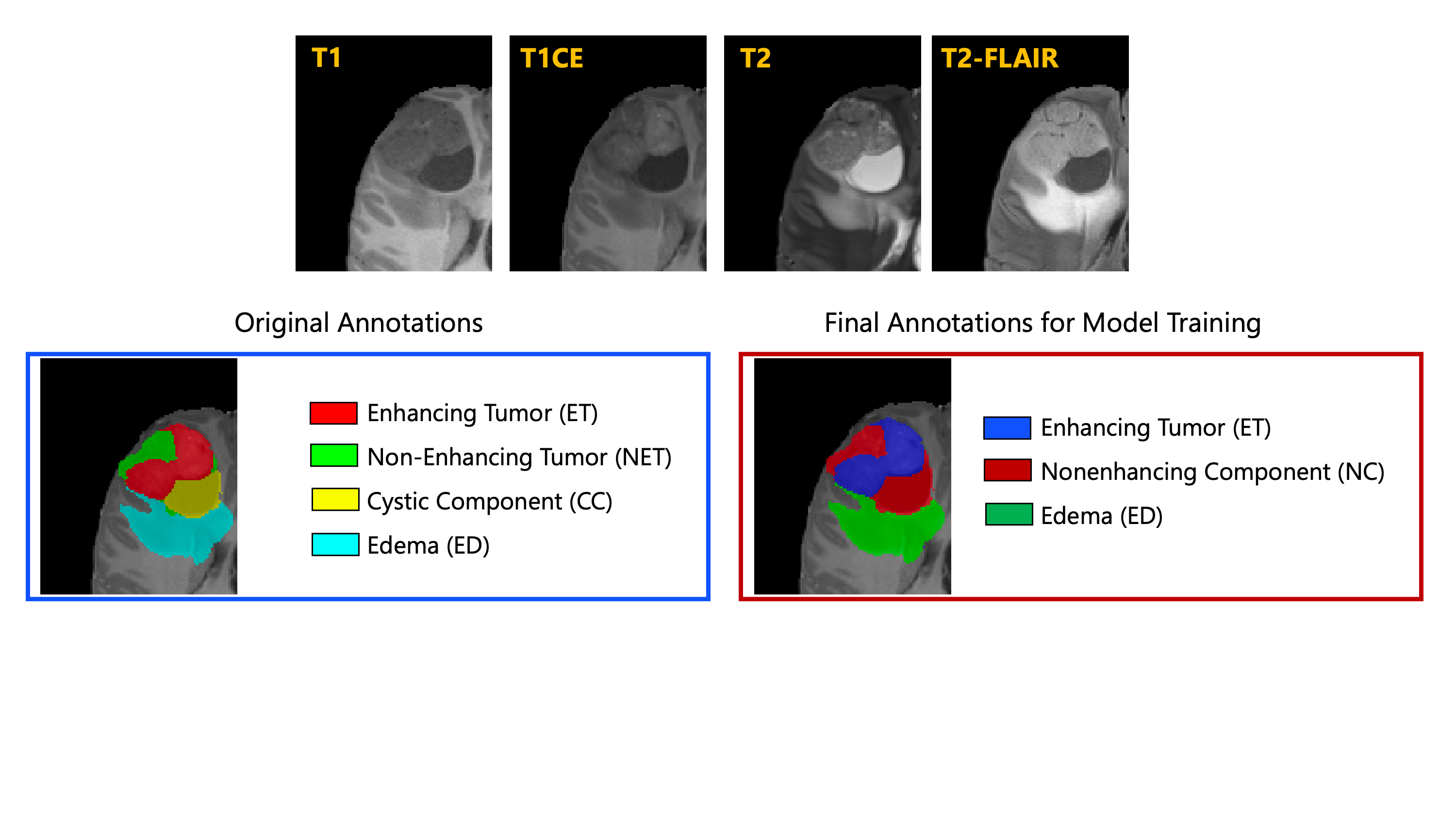}  
          \caption{\textbf{Illustrative example of tumor subregions in pediatric brain tumors.} Image panels with the annotated tumor subregions along with mpMRI structural scans (T1, T1CE, T2, and T2-FLAIR). The left-most side image on the bottom panel with the overlaid  annotations showcases the original tumor subregions, i.e., enhancing tumor (ET - red), non-enhancing tumor (NET - green), cystic component (CC - yellow), and edema (ED - teal). The image on the right hand side of the bottom panel with the overlaid annotations on the T2 sequence demonstrates the tumor subregions provided to the BraTS-PEDs 2023 participants: enhancing tumor (ET), nonenhancing component (NC), and edema (ED) \cite{fathi2023automated}.}
        \label{annotations}
    \end{figure}

    \subsubsection{Common errors of automated segmentations}

     In our experience with automated segmentation of pediatric brain tumors, some errors may be noticed:
            
            \begin{enumerate}
                \item  Peri-ventricular edema  segmented as edema
                \item  Remote areas segmented as tumor (far from actual tumor region)
                \item  Under-segmentation of cysts
                \item  Non-enhancing tumor segmented as cyst, or vice versa
            \end{enumerate}

    \subsection{Leaderboard Evaluation}
    As described earlier and exemplified in Figure 2 (bottom panel), for the BraTS-PEDs 2023 challenge, the regions to evaluate the performances are: i) the ``enhancing tumor'' (ET), ii) the ``enhancing tumor/cystic component/necrosis'' (TC)  and iii) the entire tumorous region, or the so-called ``whole tumor'' (WT).
    
    The participants are required to send the output of their methods to the evaluation platform for the scoring to occur during the training and the validation phases. At the end of the validation phase the participants are asked to identify the method they would like to evaluate in the final testing/ranking phase. The organizers will then confirm receiving the containerized method and will evaluate it on withheld testing data. The participants will be provided guidelines on the form of the container as we have done in previous years. This will enable confirmation of reproducibility, comparing these algorithms to the previous BraTS instances and comparison with results obtained by algorithms of previous years, thereby maximizing solutions in solving the problem of brain tumor segmentation. 
    
    During the training and validation phases, the participants will be able to test the functionality of their submission through three platforms:
            \begin{enumerate}
                \item Cancer Imaging Phenomics Toolkit (CaPTk [5-6], https://github.com/CBICA/CaPTk)
                \item Federated Tumor Segmentation (FeTS) Tool [7] (https://fets-ai.github.io/Front-End/)
                \item Online evaluation platform (Synapse)
            
             \end{enumerate}

    \subsection{Participation Timeline}  
        
        The challenge is composed of three main stages:
                    
            \begin{enumerate}
                \item  Training: The four MRI sequences along with the corresponding ground truth labels will be shared with participants to design and train their methods.
                \item  Validation: The validation data will be released to the participants within three weeks after the training data. The participants will not be provided with the ground truth of the validation data, but will be given the opportunity to submit multiple times to the online evaluation platforms. The participants can generate preliminary results for their trained models in unseen data and report them in their submitted short MICCAI LNCS papers , in addition to their cross-validated results on the training data, to be published in conjunction with the proceedings of the BrainLes workshop. The top-ranked participating teams in the validation phase will be invited to prepare their slides for a short oral presentation of their method during the BraTS challenge at MICCAI 2023. 
                \item  Testing/Ranking: After the participants upload their containerized method in the evaluation platforms, they will be evaluated and ranked on unseen testing data, which will not be made available. The final top-ranked participating teams will be announced at the 2023 MICCAI Annual Meeting. 
            \end{enumerate}

    \subsection{Inclusion Criteria}  
    To be completed

    \subsection{Data Availability} 
    TCIA

   \subsection{Statistical Analysis}

%% file: 3_Results.tex
\section{Results}
Overall XXX teams requested access to the BraTS-PEDs 2023 dataset, of which XXX teams participated in the validation phase. In the testing phase, nine short paper submissions were received and met the criteria for being included in the ranking.  
\subsection{Performance Evaluation}
Table XXX summarizes the methods implemented by the challenge participants along with their ranking and performance metrics. Supplementary Material XXX provides the detailed description for each tested method, submitted by the challenge participants. The best overall ranking was XXX obtained by XXX team, followed by XXX for XX team and XXX for XX team. Ranking ranged from XXX-XXX and Dice scores for the top three teams

%% file: 4_discussion.tex
\section{Discussion}

In this paper, we outlined the design of the first pediatric brain tumor segmentation (BraTS-PEDs 2023) challenge, to benchmark methods devised for segmentation of pediatric brain tumors.
In 2022, for the first time, the BraTS (continuous evaluation) challenge included testing the top-ranked methods, designed for segmentation of adult gliomas, on unseen held out data of pediatric brain tumors. The results from last year's BraTS 2022 showed that while training and optimizing the models on adult brain tumors could result in relatively high Dice scores in segmentation of the whole tumor region on mpMRI scans of pediatric brain tumors, they perform poorly on segmentation of tumor subregions (e.g. in segmentation of enhancing tumor/cystic components as a counterpart to tumor core in adult brain tumors). These results supported our hypothesis that the techniques to be applied to pediatric population need to be trained (at least partly) on curated and annotated cohort of pediatric brain tumors. However, similar to any other machine and deep learning problems, training a generalizable model mandates ample dataset processed in a standardized approach. To achieve this goal, BraTS-PEDs dataset provides the largest annotated cohort of pediatric high-grade glioma (astrocytoma, and DMG/DIPGs). We are actively working on increasing the number of subjects in this cohort to provide the community with a large dataset of these rare tumors, and to facilitate the future development of tools for computer-aided treatment planning. 
Future BraTS-PEDs challenges will include data from more institutions and will be extended to post-operative or post-treatment scans.